# First-Principles Exploration of Defect-Pairs in GaN

He Li[1#], Menglin Huang[1#] and Shiyou Chen[1,2*]

[1]*Key Laboratory of Polar Materials and Devices (MOE) and Department of Electronics, East China Normal University, Shanghai 200241, China*

[2]*Collaborative Innovation Center of Extreme Optics, Shanxi University, Taiyuan, Shanxi 030006, China*

[#]These authors contributed equally to this work. * chensy@ee.ecnu.edu.cn

Using first-principles calculations, we explored all the 21 defect-pairs in GaN and considered 6 configurations with different defect-defect distances for each defect-pair. 15 defect-pairs with short defect-defect distances are found to be stable during structural relaxation, so they can exist in the GaN lattice once formed during the irradiation of high-energy particles. 9 defect-pairs have formation energies lower than 10 eV in the neutral state. The vacancy-pair $V_N$-$V_N$ is found to have very low formation energies, as low as 0 eV in p-type and Ga-rich GaN, and act as efficient donors producing two deep donor levels, which can limit the p-type doping and minority carrier lifetime in GaN. $V_N$-$V_N$ has been overlooked in the previous study of defects in GaN. Most of these defect-pairs act as donors and produce a large number of defect levels in the band gap. Their formation energies and concentrations are sensitive to the chemical potentials of Ga and N, so their influences on the electrical and optical properties of Ga-rich and N-rich GaN after irradiation should differ significantly. These results about the defect-pairs provide fundamental data for understanding the radiation damage mechanism in GaN and simulating the defect formation and diffusion behavior under irradiation.

## 1. Introduction

As a wide band gap semiconductor, Gallium Nitride (GaN) has wide applications in light emitting diodes (LEDs), laser diodes, photodetectors and high electron mobility transistors (HEMTs) for high power, high frequency and high-temperature electronic applications.[1-4] When these devices are used for aerospace and military applications, such as satellites, communication equipment or detectors, they are usually suffering from the radiation damage caused by the proton, electron, neutron and γ-ray irradiation.[5-12] The irradiation of these high-energy particles usually causes the formation of point defects, defect-pairs, defect-complexes and even disorder regions in the semiconductor lattices.[5, 7] Although it was shown that GaN has a relatively large displacement energy $E_d$ ($E_d$ (Ga) = 18 eV, $E_d$ (N) = 22 eV) when compared to CdTe and GaAs,[6, 13-15] which means GaN can have a high radiation hardness, the irradiation of the high-energy particles may still cause the formation of various defects and change the concentration of equilibrium defects formed during growth[12]. As a result of the high energy of the irradiation particles, some high-energy defects such as the point defects that do not form during the growth, some defect-pairs or defect-complexes may be formed during the collision cascade following the primary knock-on atom (PKA) event, *e.g.*, it was shown in GaN that the concentration of (N-N)$_N$ split interstitial defects increases after the high dose of proton irradiation.[16]

As we know, there are 6 intrinsic point defects in GaN, including the vacancies $V_{Ga}$ and $V_N$, antisites $Ga_N$ and $N_{Ga}$, interstitials $Ga_i$ and $N_i$. The properties of these point defects have been well studied in the past three decades[17-21]. Besides these point defects, the 6 point defects can bind with each other to form the double-site defect-pairs and even multiple-site defect-complexes. For example, $V_{Ga}$ can bind with other defects as well as itself to form 6 defect-pairs, including $V_{Ga}$-$V_{Ga}$, $V_{Ga}$-$V_N$, $V_{Ga}$-$Ga_N$, $V_{Ga}$-$N_{Ga}$, $V_{Ga}$-$Ga_i$ and $V_{Ga}$-$N_i$. In principle, 21 types of defect-pairs can be formed, as shown in Fig. 1. As the distance and relative position between the two point defects can be different, there are a lot of possible structural configurations for these defect-pairs. These defect-pairs usually have much higher formation energies than the single point defect, so they have low equilibrium concentration in the synthesized GaN. Therefore, only a small number of defect-pairs such as $V_{Ga}$-$V_N$[22] and $V_{Ga}$-$Ga_N$[23] have been studied and their properties (formation energies and transition energy levels) are known. However, in the radiation-damaged GaN samples, these high-energy defect-pairs may have high concentrations due to the high energy injected by the irradiation particles. Unfortunately, because the fundamental properties of these defect-pairs are unknown, it is difficult to simulate the defect formation and diffusion behavior in GaN after irradiation[24] and thus difficult to predict their influences on the material properties and the device performance.

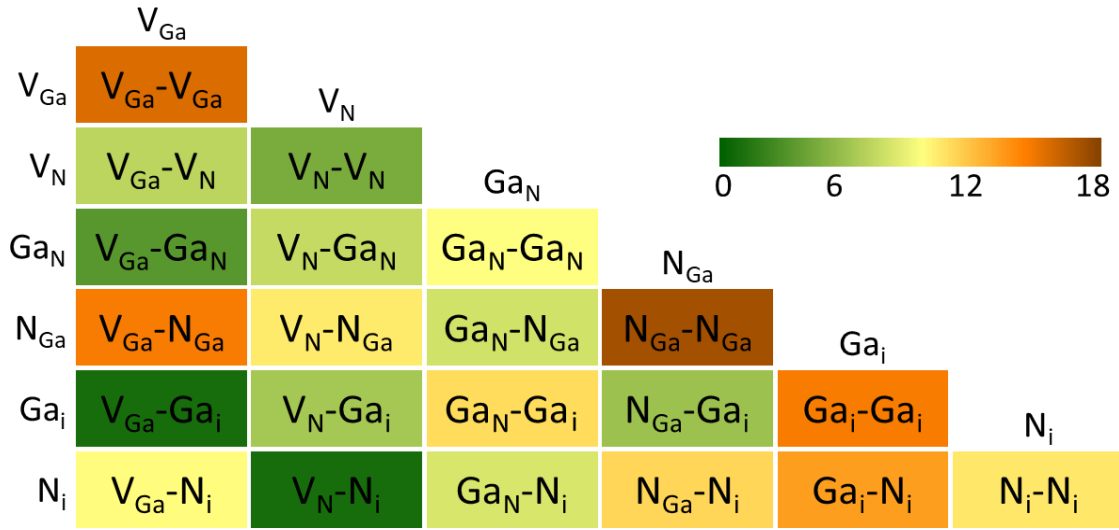

Figure 1. The 21 defect-pairs that can be formed by two point defects in GaN. The color represents their formation energies in the neutral state (in Ga-rich GaN).

In this paper we performed a systematical study on all the 21 defect-pairs in GaN through first-principles calculations of their structures, formation energies and transition energy levels. 126 structural configurations of 21 defect-pairs are considered. Most of these defect-pairs are found to be stable with short defect-defect distances and 9 defect-pairs are identified with formation energies lower than 10 eV, so they may exist with a high concentration and play important roles in the irradiated GaN. They mainly act as donors, producing many defect levels in the band gap of GaN. Their formation energies and concentrations differ significantly in the Ga-rich and N-rich GaN, indicating that the defect formation and diffusion behavior after irradiation should be very different in the Ga-rich and N-rich GaN. Among these defect-pairs, we identified one important but overlooked defect-pair, the vacancy-pair $V_N$-$V_N$, which acts as a deep donor and has very low formation energies and thus high concentration in p-type and Ga-rich GaN. Its importance in limiting the p-type doping and the minority carrier lifetime is pointed out. These results about defect-pairs are fundamental for understanding the radiation damage mechanism in GaN and also directly useful for the multiscale simulation of the defect formation and diffusion processes in the irradiated GaN, *e.g.*, using the IM3D code.

## 2. Calculation Methods

Our first-principles calculations are performed using the density functional theory and the projector-augmented wave method,[25-27] as implemented in the Vienna ab initio simulation package (VASP) code.[28] For the exchange-correlation functional, the generalized gradient approximation (GGA) in the Perdew-Burkes-Ernzerhof (PBE) form and the hybrid functional in the Heyd-Scuseria-Ernzerhof (HSE) form are adopted and compared.[29-31] The static calculations of total energies and eigenvalues are performed using the HSE functional, and the ratio of nonlocal Hartree-Fock exchange is set to 0.31, which predicts a band gap of 3.54 eV, very close to the experimental band gap of GaN.[32] We use a quasi-cubic 128-atom supercell model for calculating the defect properties, single Γ point for the Brillouin zone integration and an energy cutoff of 400 eV for the plane-wave basis set. The spin-polarization is included in all the calculations. The defect formation energy is calculated following,[17]

$$\Delta H_f(\alpha, q) = E(\alpha, q) - E(GaN) + \sum n_i \mu_i + q(E_F + E_{VBM})$$

where $E(\alpha, q)$ is the total energy of the supercell with a defect $\alpha$ in the charge state $q$, and $E(GaN)$ is the total energy of a perfect crystal GaN in the same supercell. $\mu_i$ is the chemical potential of the element $i$. $\mu_{Ga}$ and $\mu_N$ can vary in a range which is determined by the calculated formation enthalpy of GaN ($\Delta H_f(GaN)$ = -1.22 eV). $\mu_{Ga}$ can change from -1.22 eV to 0 eV (from the Ga-poor and N-rich condition to the Ga-rich and N-poor condition) with $\mu_N$ oppositely changing from 0 to -1.22 eV. $n_i$ represents the number of atoms removed from ($n_i$=+1) or added to ($n_i$=-1) the supercell in the process of forming a defect. $E_{VBM}$ is the eigenvalue of the valence band maximum (VBM) level which is aligned referenced to the electrostatic potential far from the defect site in the supercell and $E_F$ is the Fermi level referenced to the VBM level. Using the method of Makov and Payne,[33, 34] we applied the image charge corrections caused by the finite supercell size. The transition energy levels can be calculated from the formation energies of the ionized defects in different charge states.

## 3. Results and discussion

### 3.1 Structural Configurations of Defect-Pairs

In the binary compound GaN, there are 6 intrinsic point defects, including two vacancies ($V_{Ga}$, $V_N$), two interstitials ($Ga_i$, $N_i$) and two antisites ($Ga_N$, $N_{Ga}$). In principle, they can form 21 types of defect-pairs. As the two point defects are located on different sites, the formed defect-pairs can have many structural configurations. For each defect-pair, we construct 6 structural configurations with different defect-defect distances. Taking $V_N$-$V_N$ as an example, its 6 different configurations are shown in Fig. 2. The distance between the two $V_N$ increases from $(V_N-V_N)$-1 to $(V_N-V_N)$-6. For $(V_N-V_N)$-1 and $(V_N-V_N)$-2, the two $V_N$ are around the same Ga and the distances are both about 3.18 Å, however, for the other four configurations, the two $V_N$ are not around the same Ga, so the distances increase to 4.51, 5.18, 5.53 Å, as listed in Table 1.

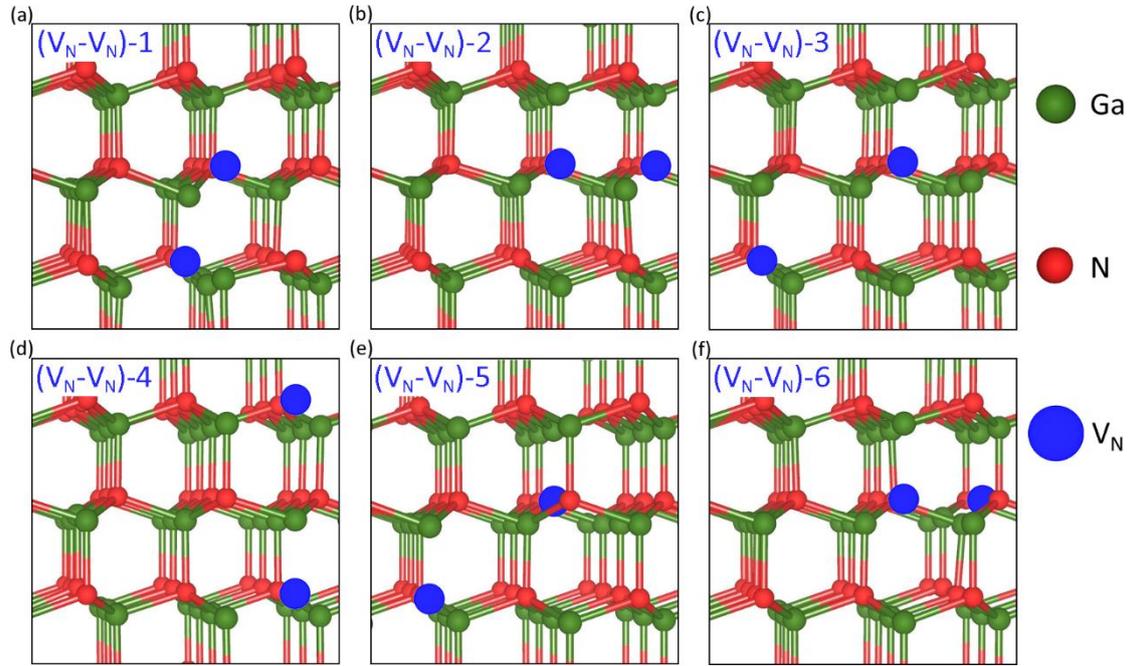

Figure 2. Six different structural configurations of the $V_N$-$V_N$ defect-pair (classified as Group-1). The blue balls show the locations of the two nitrogen vacancies ($V_N$).

As a result of the equivalency of the Ga and N sites in the wurtzite structure, the structural configurations of $V_{Ga}$-$V_{Ga}$ are same as those of $V_N$-$V_N$. Furthermore, $V_N$-$Ga_N$, $Ga_N$-$Ga_N$, $V_{Ga}$-$N_{Ga}$ and $N_{Ga}$-$N_{Ga}$ also have the same structural configurations as $V_{Ga}$-$V_{Ga}$ and $V_N$-$V_N$, because these defect-pairs are also formed on two Ga or two N sites. We classify these 6 defect-pairs as the Group-1.

In contrast, $V_{Ga}$-$V_N$, $V_{Ga}$-$Ga_N$ $V_N$-$N_{Ga}$ and $Ga_N$-$N_{Ga}$, are located on one Ga site and one N site, so they are classified as Group-2, and the distances between the two sites are listed in Table 2. The defect-pairs located on one Ga or N site and one interstitial site, including $V_{Ga}$-$Ga_i$, $V_{Ga}$-$N_i$, $V_N$-$Ga_i$, $V_N$-$N_i$, $Ga_N$-$Ga_i$, $Ga_N$-$N_i$, $N_{Ga}$-$Ga_i$ and $N_{Ga}$-$N_i$, are classified as Group-3. The three interstitial-pairs $Ga_i$-$Ga_i$, $Ga_i$-$N_i$ and $N_i$-$N_i$ are classified as Group-4.

**3.2 Defect Structures Relaxed using different Functionals**

With the structural configurations of the 21 defect-pairs, we then calculated their formation energies. The hybrid functional was shown to be more accurate in calculating the band gap of GaN and thus the formation energies and transition energy levels of defects in GaN than the GGA functional, and the calculated results agree better with the available experimental measurements.[20, 35, 36] Therefore, it would be ideal that we calculate the properties of all the defect-pairs in different configurations (21*6=126 configurations in total) using the hybrid functional. Unfortunately, the structural relaxation using the hybrid functional for all the 126 configurations of 21 defect-pairs is computationally very heavy, so we performed the structural relaxation of the defect-pairs using the computationally saving GGA-PBE functional and then performed the static calculations of total energies and eigenvalues using the HSE hybrid functional, so that the error in the calculated formation energies and transition energy levels caused by the band gap underestimation may be largely overcome and meanwhile the computational cost is affordable.

Table 1. The formation energies of two defect-pairs $V_N$-$V_N$ and $V_{Ga}$-$V_N$ in the neutral state calculated with PBE-relaxed and HSE-relaxed structures (in Ga-rich GaN). 6 structural configurations with different defect-defect distance (in the unrelaxed initial structure) are considered.

| Defect-Pair Configurations | Initial Distance (Å) | Formation Energy (eV) PBE-relaxed | Formation Energy (eV) HSE-relaxed |
|---|---|---|---|
| ($V_N$-$V_N$)-1 | 3.18 | 4.67 | 4.58 |
| ($V_N$-$V_N$)-2 | 3.19 | 4.90 | 4.81 |
| ($V_N$-$V_N$)-3 | 4.51 | 6.83 | 6.39 |
| ($V_N$-$V_N$)-4 | 5.18 | 6.72 | 6.57 |
| ($V_N$-$V_N$)-5 | 5.52 | 6.85 | 6.42 |
| ($V_N$-$V_N$)-6 | 5.53 | 6.85 | 6.41 |
| ($V_{Ga}$-$V_N$)-1 | 1.95 | 7.33 | 7.16 |
| ($V_{Ga}$-$V_N$)-2 | 1.96 | 7.37 | 7.18 |
| ($V_{Ga}$-$V_N$)-3 | 3.23 | 9.37 | 9.06 |
| ($V_{Ga}$-$V_N$)-4 | 3.74 | 9.73 | 8.87 |
| ($V_{Ga}$-$V_N$)-5 | 3.75 | 9.27 | 8.81 |
| ($V_{Ga}$-$V_N$)-6 | 4.54 | 9.37 | 8.89 |

In order to evaluate the error of our PBE+HSE method, we compared our PBE+HSE results (using GGA-PBE for structural relaxation and HSE for static calculations) to those using HSE for both the structural relaxation and static calculations for two defect-pairs $V_N$-$V_N$ and $V_{Ga}$-$V_N$. The formation energies of the two defect-pairs in 6 different configurations calculated using both the PBE-relaxed and HSE-relaxed structures are shown in Table. 1. For $V_N$-$V_N$, the shortest-distance configuration ($V_N$-$V_N$)-1 is energetically the most favorable configuration with formation energies of 4.67 eV and 4.58 eV using the PBE- and HSE-relaxed structure, respectively. $(V_N-V_N)$-2 has an energy 0.23-0.33 eV higher, while other long-distance configurations have much higher energies, at least 1.5 eV higher. The significant energy increase from the short-distance configurations to the long-distance configurations indicates that the two $V_N$ tend to bind with each other to form a short-distance defect-pair, which is independent of the specific functionals used for the structural relaxation. For $V_{Ga}-V_N$, $(V_{Ga}-V_N)$-1 and $(V_{Ga}-V_N)$-2 are energetically the most favorable and their formation energies are almost degenerate. Other long-distance configurations also have much higher energies, at least 1.6 eV higher. For both $V_N-V_N$ and $V_{Ga}-V_N$, our PBE+HSE method predicts the correct lowest-energy configurations, consistent with those predicted by the full HSE calculations. For the lowest-energy configurations, the formation energy differences between the PBE-relaxed and HSE-relaxed results are less than 0.2 eV, while for the higher-energy configurations, the formation energy differences can be as large as 0.9 eV, indicating that our calculations using the PBE-relaxed structures can predict the formation energies with an error less than 0.9 eV. Compared to the absolute value of the formation energy around 9 eV, the error is about 10%. This error is acceptable for the high-throughput exploration of defect-pairs that may form in the irradiated GaN with radiation damage.

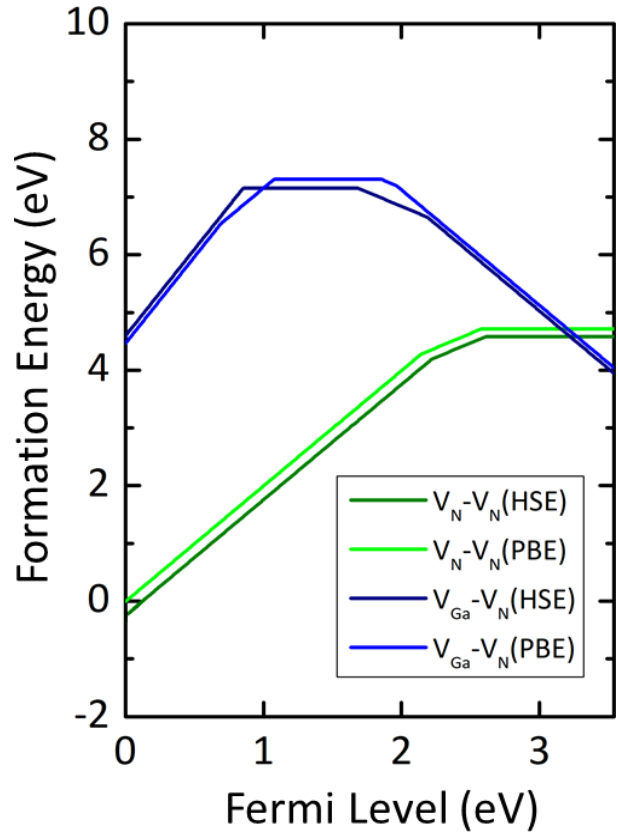

Figure 3. The formation energies of two defect-pairs $V_N-V_N$ and $V_{Ga}-V_N$ in different

charge states as functions of the Fermi level (in Ga-rich GaN), calculated with PBE-relaxed and HSE-relaxed structures.

For the lowest-energy configurations of $V_N$-$V_N$ and $V_{Ga}$-$V_N$, we further calculated the formation energies in different charge states which change with the Fermi level, as shown in Fig. 3. The comparison shows that the results for the charged defect-pairs of our PBE+HSE method are also in good agreement with those of the full HSE method (Note that the PBE- and HSE-relaxed structures differ slightly for $V_{Ga}$-$V_N$, resulting in the different transition energy levels). Therefore, in the following, we will use the PBE+HSE (PBE relaxation and HSE static calculation) method to calculate the formation energies and transition energy levels of 21 intrinsic defect-pairs with different structural configurations.

### 3.3 Unstable Defect-Pairs relaxing into Point Defects

The calculated formation energies of 21 intrinsic defect-pairs in the neutral charge state are shown in Table 2. According to their structures, the 21 defect-pairs are classified into four groups, as mentioned in Section 3.1. For each defect-pair, 6 structural configurations with different distances between two defect sites are considered, and the distances are listed in Table 2. Both the Ga-rich and N-rich conditions are considered. For the defect-pairs with formation energies lower than 10 eV in the neutral state, we further calculated the formation energies in different charge states, as shown in Fig. 4 where the formation energies of single-site point defects are also plotted for comparison.

Table 2. The calculated formation energies (in eV) of 21 defect-pairs in Ga-rich and N-rich GaN. For each defect-pair, 6 different structural configurations are considered and the distances (in Å) between the two defect sites are listed (before structural relaxation). The lowest-energy configurations are shown in bold. The symbol * means that the defect-pair becomes a point defect or annihilated after structural relaxation.

| **Group-1** | | | | | | | |
|---|---|---|---|---|---|---|---|
| Distance | | 3.18 | 3.19 | 4.51 | 5.18 | 5.52 | 5.53 |
| Ga rich | $V_{Ga}$-$V_{Ga}$ | **15.81** | 15.84 | 15.96 | 15.89 | 15.89 | 15.87 |
| | $V_{Ga}$-$N_{Ga}$ | 15.82 | **15.31** | 20.95 | 17.98 | 18.46 | 17.61 |
| | $V_N$-$V_N$ | **4.67** | 4.90 | 6.83 | 6.72 | 6.85 | 6.85 |
| | $V_N$-$Ga_N$ | 7.67 | **7.53** | 8.62 | 8.06 | 8.63 | 8.57 |
| | $Ga_N$-$Ga_N$ | **9.73** | 9.74 | 12.57 | 12.13 | 12.60 | 12.64 |
| | $N_{Ga}$-$N_{Ga}$ | **17.40** | 17.49 | 20.32 | 20.11 | 20.17 | 20.26 |
| N rich | $V_{Ga}$-$V_{Ga}$ | **13.37** | 13.40 | 13.53 | 13.45 | 13.45 | 13.43 |
|  | $V_{Ga}$-$N_{Ga}$ | 12.17 | **11.66** | 17.30 | 14.32 | 14.81 | 13.96 |
|  | $V_N$-$V_N$ | **7.10** | 7.32 | 9.26 | 9.15 | 9.27 | 9.28 |
|  | $V_N$-$Ga_N$ | 11.32 | **11.19** | 12.27 | 11.72 | 12.28 | 12.22 |
|  | $Ga_N$-$Ga_N$ | **14.60** | 14.61 | 17.44 | 17.00 | 17.47 | 17.51 |
|  | $N_{Ga}$-$N_{Ga}$ | **12.53** | 12.62 | 15.45 | 15.24 | 15.30 | 15.39 |
| **Group-2** | | | | | | | |
| Distance |  | 1.95 | 1.96 | 3.23 | 3.74 | 3.74 | 4.54 |
| Ga rich | $V_{Ga}$-$V_N$ | **7.33** | 7.37 | 9.37 | 9.73 | 9.27 | 9.37 |
|  | $V_{Ga}$-$Ga_N$ | 3.28* | **3.26*** | 3.30* | 3.28* | **3.26*** | 3.30* |
|  | $V_N$-$N_{Ga}$ | 10.47 | **10.42** | 11.64 | 11.74 | 11.99 | 11.63 |
|  | $Ga_N$-$N_{Ga}$ | **7.99** | 8.62 | 16.50 | 16.35 | 16.02 | 15.42 |
| N rich | $V_{Ga}$-$V_N$ | **7.33** | 7.37 | 9.37 | 9.73 | 9.27 | 9.37 |
|  | $V_{Ga}$-$Ga_N$ | 4.50* | **4.47*** | 4.52* | 4.50* | **4.47*** | 4.52* |
|  | $V_N$-$N_{Ga}$ | 9.25 | **9.20** | 10.42 | 10.52 | 10.78 | 10.41 |
|  | $Ga_N$-$N_{Ga}$ | **7.99** | 8.62 | 16.50 | 16.35 | 16.02 | 15.42 |
| **Group-3** | | | | | | | |
| Distance |  | 0.97 | 1.02 | 1.57 | 1.86 | 3.69 | 4.87 |
| Ga rich | $V_{Ga}$-$Ga_i$ | 0.00* | 0.00* | 0.00* | 0.00* | 0.00* | 0.00* |
|  | $V_{Ga}$-$N_i$ | 10.40 | 10.44 | **9.92** | 9.93 | 10.64 | 10.61 |
|  | $V_N$-$Ga_i$ | 6.52* | 6.56* | **6.27*** | 6.53* | 11.36 | 11.60 |
|  | $V_N$-$N_i$ | 0.00* | 0.00* | 0.00* | 0.00* | 7.66 | 8.42 |
|  | $Ga_N$-$Ga_i$ | **11.16** | 11.69 | 12.35 | 11.17 | 12.82 | 14.10 |
|  | $Ga_N$-$N_i$ | **8.24*** | 9.23* | 8.75* | 9.02* | 10.81 | 12.73 |
|  | $N_{Ga}$-$Ga_i$ | 5.98* | 6.10* | 6.29* | 6.26* | 17.41 | **5.98*** |
|  | $N_{Ga}$-$N_i$ | **11.37** | 15.87 | 13.44 | 11.46 | 11.81 | 15.10 |
| N rich | $V_{Ga}$-$Ga_i$ | 0.00* | 0.00* | 0.00* | 0.00* | 0.00* | 0.00* |
|  | $V_{Ga}$-$N_i$ | 7.97 | 8.00 | **7.48** | 7.50 | 8.21 | 8.18 |
|  | $V_N$-$Ga_i$ | 8.95* | 9.00* | **8.71*** | 8.96* | 13.80 | 14.03 |
|  | $V_N$-$N_i$ | 0.00* | 0.00* | 0.00* | 0.00* | 7.66 | 8.42 |
|  | $Ga_N$-$Ga_i$ | **14.81** | 15.34 | 16.00 | 14.82 | 16.47 | 17.75 |
|  | $Ga_N$-$N_i$ | **9.45*** | 10.45* | 9.97* | 10.24* | 12.03* | 13.95 |
|  | $N_{Ga}$-$Ga_i$ | 4.76* | 4.88* | 5.07* | 5.04* | 16.19 | **4.76*** |
|  | $N_{Ga}$-$N_i$ | **7.72** | 12.22 | 9.79 | 7.80 | 8.15 | 11.45 |
| **Group-4** | | | | | | | |
| Distance |  | 1.62 | 1.98 | 2.62 | 2.83 | 3.20 | 3.82 |
| Ga rich | $Ga_i$-$Ga_i$ | **14.92** | 14.93 | 15.94 | 14.64 | **14.92** | 16.13 |
|  | $Ga_i$-$N_i$ | 14.08 | **13.51** | 13.54 | 14.67 | 14.07 | 27.75 |
|  | $N_i$-$N_i$ | 11.14 | **10.56** | 10.85 | 12.28 | 18.05 | 15.49 |
| N rich | $Ga_i$-$Ga_i$ | **17.36** | 17.36 | 18.38 | 17.07 | **17.36** | 18.57 |
|  | $Ga_i$-$N_i$ | 14.08 | **13.51** | 13.54 | 14.67 | 14.07 | 27.75 |
|  | $N_i$-$N_i$ | 8.70 | **8.13** | 8.41 | 9.84 | 15.62 | 13.06 |

Two defect-pairs $V_{Ga}$-$Ga_i$ and $V_N$-$N_i$ (Frenkel defect pair) are found to be unstable and annihilated during the structural relaxation. The structural relaxation of $V_{Ga}$-$Ga_i$ shows that the interstitial Ga moves back to the Ga vacancy and fills the vacancy, causing the annihilation of the Frenkel defect pair, so the calculated formation energy becomes zero in Table 2 (the crystalline lattice becomes defect-free). For $V_{Ga}$-$Ga_i$, no matter how large the distance between the vacancy and the interstitial is, the Ga interstitial can relax back to the vacancy during the structural relaxation, so the formation energies of all the configurations are 0 in Table 2. Therefore, we can conclude that $V_{Ga}$-$Ga_i$ cannot exist as a stable Frenkel defect pair. Four short-distance configurations (shorter than 2 Å) of $V_N$-$N_i$ cannot exist as stable Frenkel defect pairs either, because the interstitial N atom also relaxes back to the N vacancy site. However, when the distance between $V_N$ and $N_i$ increases to 3.69 Å and 4.87 Å, the formation energies of $V_N$-$N_i$ increase to 7.66 eV and 8.42 eV, respectively, indicating that there is a barrier preventing the interstitial N from relaxing back to the vacancy site, so these $V_N$-$N_i$ configurations are stable Frenkel defect pairs although the distance is quite large. The origin is that the N anion is relatively small, so it can stay on the interstitial site as a metastable state when the interstitial is far from the vacancy. The case is different from that for the large Ga cation, which cannot stay on the interstitial site as a metastable state and there is no barrier preventing it from relaxing back to the vacancy site.

Different from $V_{Ga}$-$Ga_i$ and $V_N$-$N_i$ which are annihilated after the structural relaxation, $V_{Ga}$-$Ga_N$ becomes a point defect $V_N$ after the structural relaxation, because the antisite Ga moves back to the Ga vacancy. Similarly, the short-distance configuration of $V_N$-$Ga_i$ becomes a point defect $Ga_N$ because the interstitial Ga moves to occupy the N vacancy during the structural relaxation. As shown in Fig. 4(a) and 4(b), $V_{Ga}$-$Ga_N$ and $V_N$-$Ga_i$ have the same formation energies as $V_N$ and $Ga_N$, respectively. Thus, the formation energies of $V_{Ga}$-$Ga_N$ and $V_N$-$Ga_i$ are plotted in dashed lines. Due to its large radius, the Ga atom on the N site or the interstitial site tends to relax to the vacancy site, so $V_{Ga}$-$Ga_N$ and the short-distance $V_N$-$Ga_i$ becomes point defects and cannot exist as stable defect-pairs. When the distance between $V_N$ and $Ga_i$ is larger than 3.5 Å, the Ga relaxation is prevented by a barrier, making the long-distance $V_N$-$Ga_i$ exist as a stable defect-pair.

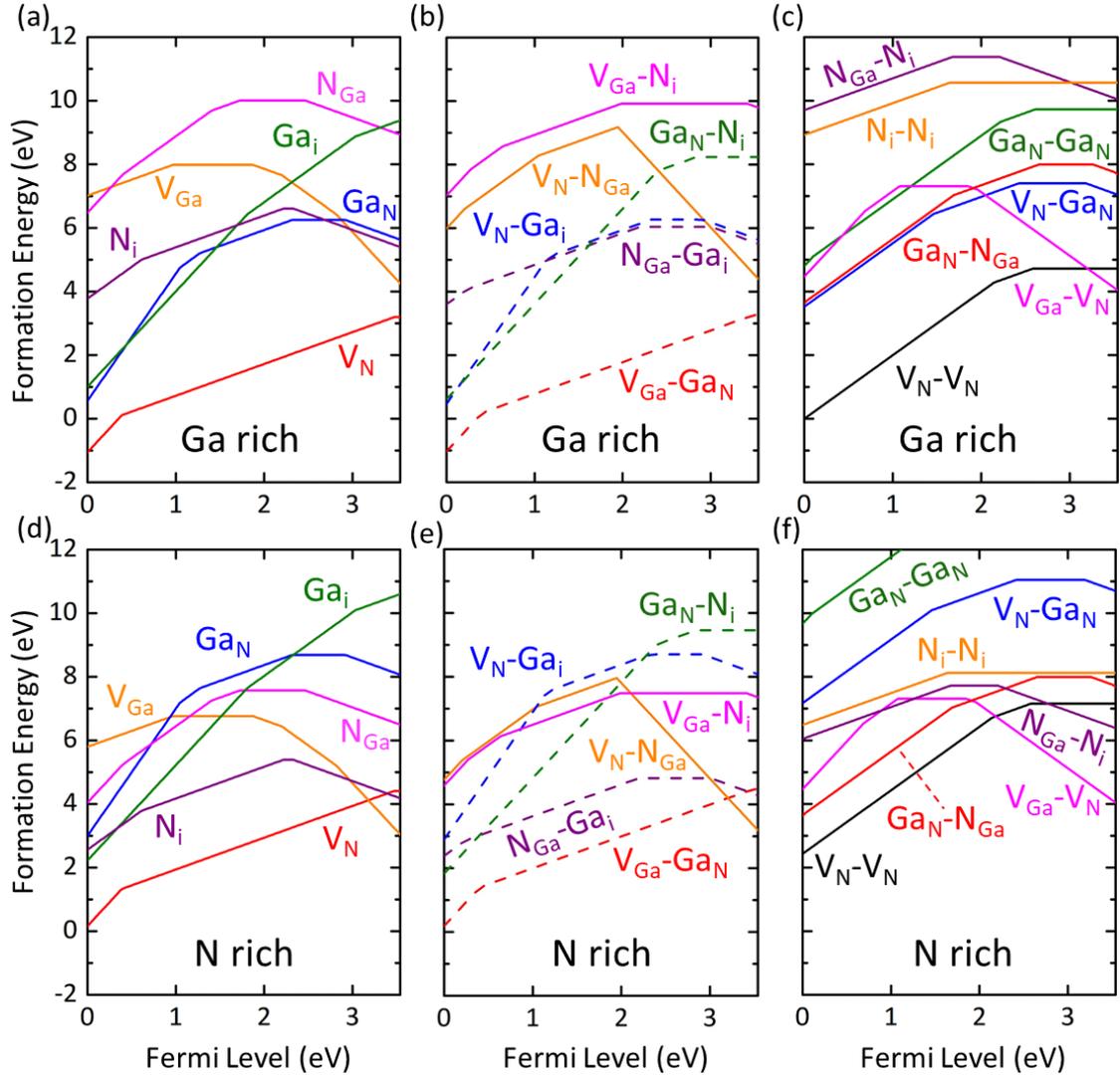

Figure 4. The formation energies of, (a) 6 point defects, (b) 6 defect-pairs that have the same component atoms as point defects, (c) 7 defect-pairs that cannot relax to point defects, as functions of Fermi level in Ga-rich GaN. (d-f) shows the corresponding results in N-rich GaN.

In contrast to $V_{Ga}$-$Ga_N$ and $V_N$-$Ga_i$, $V_N$-$N_{Ga}$ and $V_{Ga}$-$N_i$ can exist as stable defect-pairs. The relaxation of $V_N$-$N_{Ga}$ shows that the antisite N does not move to the N vacancy site. The relaxed $V_{Ga}$-$N_i$ is also structurally different from $V_{Ga}$. Thus, the formation energies of $V_N$-$N_{Ga}$ and $V_{Ga}$-$N_i$ differ from those of $V_{Ga}$ and $N_{Ga}$, respectively, as shown in Fig. 4. The reason is that the small N atom can be accommodated by the Ga site and the interstitial site with a smaller energy cost.

For the two antisite-interstitial defect-pairs $Ga_N$-$N_i$ and $N_{Ga}$-$Ga_i$, they can relax to the point defects $Ga_i$ and $N_i$, respectively. But our structural relaxation shows that the relaxed structures of their 5 low-energy configurations are not exactly same as those of

$Ga_i$ and $N_i$, so these structures can be considered as the new structures of $Ga_i$ and $N_i$ point defects. Therefore, the formation energies of $Ga_N$-$N_i$ and $N_{Ga}$-$Ga_i$ in Fig. 4(b) and 4(e) are different from those of $Ga_i$ and $N_i$ in Fig. 4(a) and 4(d).

### 3.3 Properties of Stable Defect-Pairs

Except the two defect-pairs that are annihilated and four defect-pairs that relax to point defects after structural relaxation, the other 15 defect-pairs are stable during the structural relaxation. As listed in Table 2, the lowest-energy configurations are generally those with short distances, and the longer-distance configurations have higher energies, indicating that the defect-pairs are stable because there is an energy barrier preventing them from separating into two isolated point defects. It should be noted that our criterion for stable defect-pairs here are different from the common criterion for stable defect-pairs. The common criterion for the thermodynamically stable defect-pairs is that the binding energy of the two point defects forming the defect-pair is negative, *i.e.*, the formation energy of the defect-pair is lower than the sum of the formation energies of two isolated point defects. In the equilibrium state, this criterion is valid for the defect formation because the defect concentration is determined by its formation energy, and a high-energy defect-pair with a positive binding energy tends to separate into two isolated point defects, which is thermodynamically favorable. However, under the irradiation of high-energy particles, the semiconductors are in the non-equilibrium state and some high-energy defect-pairs may be formed during the collision cascade process following the primary knock-on atom (PKA) event. If these defect-pairs are not annihilated or relax to point defects after structural relaxation, and their short-distance configurations have lower energies than longer-distance configurations (producing an energy barrier), these defect-pairs can exist in the lattice for a long period and be stable dynamically, even though they are not at the thermodynamic ground state. For studying the defects in the radiation-damaged semiconductors, we adopt this criterion for the stability of defect-pairs.

Although the formation and concentration of defects are not solely determined by the formation energies in the non-equilibrium state, the formation energies are still important because they influence the defect formation and diffusion processes directly under irradiation and the values are necessary for the multiscale simulation of these processes[24]. There are 15 defect-pairs that have formation energies lower than 10 eV in Ga-rich or N-rich GaN samples. These defect-pairs should form more easily than other defect-pairs, and may play important roles in influencing the properties of the radiation-damaged GaN. In order to provide more information for the future characterization studies on these defect-pairs, we also calculated their formation energies at different charge states (Fig. 4), from which the transition energy levels can be derived, as shown in Fig. 5.

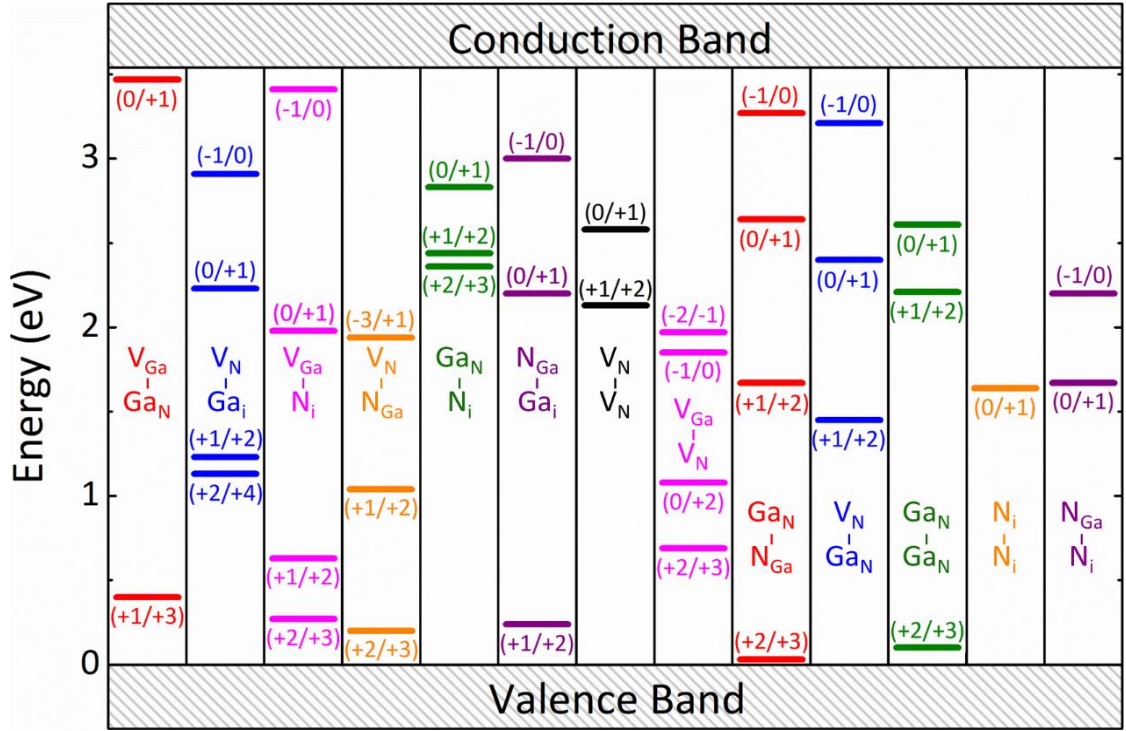

Figure 5. The transition energy levels of the 13 defect-pairs in the band gap of GaN.

In Fig. 4(b) and 4(e), the formation energies of the 6 defect-pairs that have the same component atoms as point defects are shown. As we see, the results in Fig. 4(b) are similar to those in Fig. 4(a), and Fig. 4(e) similar to Fig. 4(d). In Ga-rich GaN, the stable defect-pair $V_N$-$N_{Ga}$ has a formation energy around 4 eV in n-type GaN (when the Fermi level is close to CBM level) and a formation energy around 6 eV in p-type GaN (the Fermi level is close to VBM level). In N-rich GaN, the values are decreased to 3 eV and 5 eV respectively. Another stable defect-pair $V_{Ga}$-$N_i$ has a higher formation energy, above 7 eV in Ga-rich GaN and above 4.5 eV in N-rich GaN. The two stable defect-pairs produce 7 transition energy levels in the band gap, which are quite different from the levels of the point defects $V_{Ga}$ and $N_{Ga}$ with the same atom composition.

In Fig. 4(c) and 4(f), the formation energies of the 7 defect-pairs that cannot relax to any point defects are shown. In Ga-rich GaN, most of the defect-pairs have formation energies in the range 4-10 eV, but one defect-pair $V_N$-$V_N$, has a very low formation energy, as low as 0 eV in p-type GaN. In N-rich GaN, most of the defect-pairs have formation energies in the range 4-8 eV, and $V_N$-$V_N$ has a formation energy as low as 2.5 eV in p-type GaN. The low formation energies of these defect-pairs indicate that they may form and exist with a high concentration in the irradiated GaN because the high-energy particles and the primary knock-on atoms (PKA) can carry much higher energies and collide with other atoms to form these defect-pairs.

Among these low-energy defect-pairs, $V_N$-$V_N$ has the lowest energy and should be the most favorable defect-pair in GaN. Its two short-distance configurations both have low formation energies, but ($V_N$-$V_N$)-1 with the two vacancies in two different (0001) planes is more stable than ($V_N$-$V_N$)-2 with the two vacancies in the same (0001) plane, and the energy difference is about 0.23 eV. In the neutral state, the formation energy of $V_N$-$V_N$ is 4.72 eV in Ga-rich GaN and 7.16 eV in N-rich GaN. It is a donor defect, so it can be ionized to +1 and +2 charged states, then its formation energy can be decreased as the Fermi level is decreased to close to VBM in p-type GaN. In Ga-rich and p-type GaN, the +2 charged $V_N$-$V_N$ has a formation energy around 0 eV, indicating that this defect-pair can have a very high concentration if Ga-rich GaN samples are doped into highly p-type. In N-rich GaN, the +2 charged $V_N$-$V_N$ has a formation energy higher than 2 eV, but it's still the lowest among all the defect-pairs. The very low formation energy and thus the easy formation of $V_N$-$V_N$ defect-pairs have never been reported as far as we know, despite hundreds of papers about the defects in GaN have been published since 1990s[17-21]. This is important for not only the study of defects in radiation-damaged GaN samples but also for the study on the properties of un-irradiated GaN, *e.g.*, understanding its p-type doping limit and the optical and electrical characterization of Ga-rich GaN. As shown in Fig. 5, the defect-pair $V_N$-$V_N$ is a donor and produces two transition energy levels in the band gap, *i.e.*, the (+1/+2) level at 2.13 eV and the (0/+1) level at 2.58 eV above the VBM level. These two deep donor levels may cause the non-radiative recombination of electron-hole pairs and thus impose a limit to the minority carrier lifetime in Ga-rich GaN if it is doped into highly p-type. We expect that the two levels may be observed by the photoluminescence and deep-level transient spectroscopy experiments in Mg-doped and Ga-rich GaN and call for experimental confirmation.

The vacancy-pair, $V_{Ga}$-$V_N$, usually known as the divacancy defect in GaN[23], has a formation energy of 7.31 eV in the neutral state and as low as 4 eV in n-type GaN, and its formation energies are independent of Ga-rich or N-rich conditions. It acts as a bipolar defect, *i.e.*, it can be ionized to the +2 and +3 charge states in p-type GaN while can also be ionized to the -1 and -2 charge states in n-type GaN. In p-type GaN, its formation energy is higher than that of $V_N$-$V_N$, but in n-type GaN, its formation energy can be the lowest among all the defect-pairs, indicating that $V_{Ga}$-$V_N$ should be the most important defect-pair in n-type GaN after irradiation. It produces two donor levels, (+2/+3) at 0.69 eV and (0/+2) at 1.08 eV, and two acceptor levels (-1/0) at 1.85 eV and (-2/-1) at 1.97 eV above VBM level. These results about the divacancy $V_{Ga}$-$V_N$ are in good agreement with the results of Diallo *et al*[23].

Another important but overlooked defect-pair for the defect diffusion in GaN is the antisite-pair $Ga_N$-$N_{Ga}$. We expected that this antisite-pair may be a charge-compensated defect and thus does not produce any transition energy levels in the band gap. However, our calculations showed that it is also an efficient donor, producing three donor levels in the band gap, as shown in Fig. 5. Its formation energy can be lower than 4 eV in p-type GaN, and the value is also independent of the Ga-rich or N-rich condition. Its formation energy is only higher than that of $V_N$-$V_N$ but lower than other defect-pairs, indicating that its equilibrium concentration can be the second highest in the p-type GaN after irradiation. Our literature search shows that the properties of this antisite-pair had been seldom reported, and its importance had been overlooked. As an antisite-pair, $Ga_N$-$N_{Ga}$ can be the intermediate state of the inter-diffusion of Ga cations and N anions after the irradiation injects high local energy in a certain region of the GaN lattice. Our calculated transition energy levels may be used to explain the photoluminescence experiments of the radiation-damaged GaN.

As shown in Fig. 4(c) and 4(f), most of the 7 defect-pairs tend to act donors. As the Fermi level is near the VBM level, they are ionized to the positive charge states and their formation energies are low, and as the Fermi level shifts up, their formation energies increase and they become unionized and even negatively charged. Therefore, a large number of donor levels are shown in Fig. 5. Only $V_{Ga}$-$V_N$ and $N_{Ga}$-$N_i$ produce two deep acceptor levels, and can be ionized to the positive charge states with obviously lower formation energies in the n-type GaN.

Comparing the Ga-rich and N-rich conditions, we can see that two $Ga_N$-related defect-pairs, $V_N$-$Ga_N$ and $Ga_N$-$Ga_N$ have formation energies lower than 8 eV in Ga-rich GaN while much higher energies above 10 eV in N-rich GaN. In contrast, two $N_i$-related defect-pairs, $N_{Ga}$-$N_i$ and $N_i$-$N_i$, have formation energies lower than 8 eV in N-rich GaN while much higher formation energies in Ga-rich GaN. The significant differences in the formation energies and thus the concentrations of these defect-pairs indicates that the radiation-damage effects in Ga-rich and N-rich GaN samples should be very different, and a comparison study of Ga-rich and N-rich GaN samples is necessary for studying the effects of various radiations on GaN devices.

## 4. Conclusions

The 21 defect-pairs formed by two point defects in GaN are studied using the first-principles calculations. For each defect-pair, 6 structural configurations with different defect-defect distances are considered, and 126 structural configurations of 21 defect-pairs are studied in total. The calculations showed: (i) after structural relaxation, 2 defect-pairs $V_{Ga}$-$Ga_i$ and $V_N$-$N_i$ (Frenkel defect pair) are annihilated and 4 defect-pairs $V_{Ga}$-$Ga_N$ and $V_N$-$Ga_i$, $Ga_N$-$N_i$ and $N_{Ga}$-$Ga_i$ become point defects, while the other 15 defect-pairs remain as defect-pairs with short defect-defect distances; (ii) most of these defect-pairs have lower formation energies when the defect-defect distances are small, indicating that there is an energy barrier preventing the separation of the defect-pairs into two isolated point defects and the defect-pairs can be stable once formed during the irradiation of high-energy particles; (iii) 9 stable defect-pairs have their formation energies lower than 10 eV in the neutral charge state, so they form and play important role in influencing the properties of radiation-damaged GaN; (iv) the vacancy-pair $V_N$-$V_N$ is found to have very low formation energies and thus high concentration, especially in p-type and Ga-rich GaN. It can be an important limiting defect to the p-type doping and also causes serious non-radiative recombination because it produces deep donor levels. The high concentration and importance of $V_N$-$V_N$ in p-type and Ga-rich GaN has been overlooked in the past three decades. (v) most of these defect-pairs act as donors, and their formation energies fall in the range 4-8 eV and differ significantly in the Ga-rich and N-rich GaN, indicating that the defect formation and diffusion behavior in the Ga-rich and N-rich GaN after irradiation should be quite different and a comparison study is necessary. Our calculated properties of these defect-pairs provide fundamental data for the simulation of defect formation and diffusion in the irradiated GaN. Further study on the important defect-pairs that we identified, including $V_N$-$V_N$ in Ga-rich and p-type GaN and $Ga_N$-$N_{Ga}$ in the radiation-damaged GaN, are called for.

**Acknowledgements**

The authors thank Profs. Yongsheng Zhang and Yonggang Li for telling us the importance of defect-pairs in the simulation of radiation damage. This work was supported by the Science Challenge Project (TZ2018004), National Natural Science Foundation of China (NSFC) under grant Nos. 61722402 and 91833302, National Key Research and Development Program of China (2016YFB0700700), Shanghai Academic/Technology Research Leader (19XD1421300), Fok Ying Tung Education Foundation (161060) and the Fundamental Research Funds for the Central Universities.